\documentclass{aa}  
\usepackage{color}
\usepackage{graphicx}
\usepackage{lscape}
\usepackage{natbib}
\usepackage[varg]{txfonts}
\usepackage{url}
\usepackage{xspace}
\bibpunct{(}{)}{;}{a}{}{,}

\newcommand{\Teff}{$T\mathrm{\hspace*{-0.4ex}_{eff}}$\,}
\newcommand{\logg}{$\log\,g$\hspace*{0.5ex}}

\def\kpd{KPD\,0005+5106}
\def\hh{H1504+65}
\begin{document}

\title{Hubble Space Telescope ultraviolet spectroscopy of the hottest known helium-rich pre-white dwarf \kpd}

\author{K\@. Werner \and T\@. Rauch}

\institute{Institute for Astronomy and Astrophysics, Kepler Center for Astro and
Particle Physics, Eberhard Karls University, Sand~1, 72076
T\"ubingen, Germany\\ \email{werner@astro.uni-tuebingen.de}}

\date{Received 19 August 2015 / Accepted 28 September 2015}

\authorrunning{K. Werner \& T. Rauch}
\titlerunning{HST/STIS ultraviolet spectroscopy of \kpd}

\abstract{We present a model-atmosphere analysis of the ultraviolet
  echelle spectra of \kpd\ taken with the Space Telescope Imaging
  Spectrograph aboard the Hubble Space Telescope. The star is the
  hottest known pre-white dwarf (\Teff = $200\,000\pm20\,000$\,K,
  \logg = $6.7\pm0.3$ (cm/s$^2$). Its atmosphere is composed of helium
  with trace amounts of metals. It is of the so-called O(He) spectral
  type that comprises very hot helium-rich pre-white dwarfs whose
  origin is debated. From neon and silicon ionisation balances, we
  derive tighter constraints on the effective temperature
  ($195\,000\pm15\,000$\,K) and improve previous abundance
  determinations of these elements. We confirm the idea that \kpd\ is
  the descendant of an R Coronae Borealis (RCB) star, so is the
  outcome of a binary-white-dwarf merger. We discuss the relation of
  \kpd\ to other O(He) and RCB stars.}


\maketitle
%

\section{Introduction}
\label{intro}

\kpd\ is the hottest known pre-white dwarf star. It has a
helium-dominated atmosphere with trace metals and no detectable
hydrogen \citep[H $<$ 2.5\,\% mass fraction;][]{2010A&A...524A...9W}.
It was originally classified as a hot DO white dwarf (WD) with an
effective temperature of \Teff = 120\,000\,K and a surface gravity of
\logg = 7 [cm/s$^2$] \citep{1994A&A...284..907W}, however, it was
realized later that the star is much hotter and has a lower surface
gravity (\Teff = $200\,000 \pm 20\,000$\,K, \logg = $6.7\pm0.3$),
i.e., higher luminosity, meaning that it actually is still
helium-shell burning \citep{2007A&A...474..591W}. Consequently, it
must be classified as an O(He) star, a designation introduced by
\cite{1991IAUS..145..375M} for helium-dominated pre-white dwarfs with
spectra characterized by absorption lines from ionized helium.

The origin of the O(He) stars is unclear and currently debated
\citep[e.g., ][]{2014A&A...566A.116R}. The majority of the pre-white
dwarfs are rich in  hydrogen and compatible with canonical stellar
evolution theory \citep[e.g., ][]{1995PhR...250....2I}. A large
number are, however, hydrogen-deficient, and most of these objects are
PG\,1159 stars that are also helium-rich but, in contrast to the
O(He) stars, also strongly enriched in carbon and oxygen. The PG\,1159
surface chemistry is explained by a late helium-shell flash that
consumes the hydrogen envelope and dredges up helium, enriched by
He-burning ashes from the stellar core \citep{1983ApJ...264..605I,
  1999A&A...349L...5H, 2006PASP..118..183W}. It has been argued that the
O(He) stars represent a distinct post-AGB sequence, possibly initiated
by binary WD mergers, and it was speculated that they are descendants
of the R~Coronae Borealis (RCB) stars. This evolutionary link was
invoked particularly for \kpd\ because its trace element abundances
are rather similar \citep{2008ASPC..391..135R,2010A&A...524A...9W}. 

RCB stars are hydrogen-deficient supergiants
\citep{1996PASP..108..225C} with \Teff = 4000 -- 8000\,K and \logg =
0.5 -- 1.5. There is now general consensus that RCB stars have a
binary WD merger origin \citep{1984ApJ...277..355W}, based on evidence
from evolution timescales, pulsation masses, and surface element
abundances \citep{2002MNRAS.333..121S, 2006ApJ...638..454P,
  2007ApJ...662.1220C}.  Closely related are the Extreme Helium stars
(EHe), which are hotter, early-type (A and B) supergiants with similar
photospheric composition, and they are assumed descendants of the RCB
stars \citep[e.g.,][]{2011MNRAS.414.3599J}.

The abundance pattern of \kpd\ was derived by the
analysis of optical but mainly far-ultraviolet spectra taken with the
Far-Ultraviolet Spectroscopic Explorer (FUSE), covering the wavelength
range 912--1180\,\AA\ \citep{2008ASPC..391..135R,
  2010A&A...524A...9W}. In this paper, we present new
ultraviolet (UV) spectra taken with the Hubble Space Telescope (HST),
stretching the observed spectral range up to the optical. Our
observations were intended to independently check for the effective
temperature determination and to improve the determination of trace
element abundances to better constrain the relation of \kpd\ to the
RCB stars and to shed more light on the evolutionary status of the
O(He) stars.

We begin with a description of the observations
(Sect.\,\ref{sect:observations}) and continue with a delineation of
our model atmospheres and model atoms (Sect.\,\ref{sect:models})
utilized for the spectral analysis. In Sect.\,\ref{sect:results} we
present in detail the line identifications and line fitting
procedure. Finally, the results are summarized and discussed in the
context of RCB and O(He) stars in Sect.\,\ref{sect:discussion}.

\begin{figure}[t]
 \centering  \includegraphics[width=1.0\columnwidth]{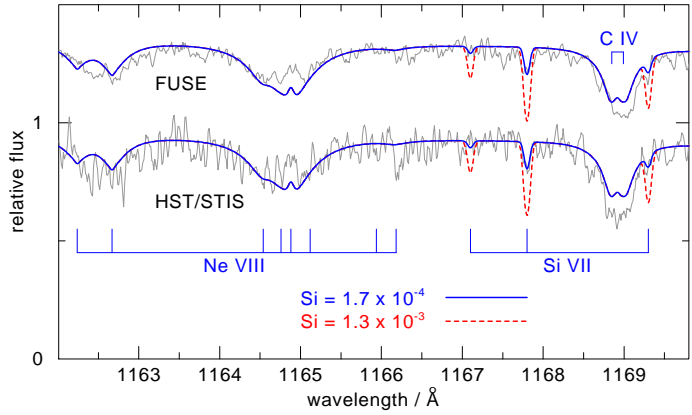}
  \caption{Detail of the HST/STIS spectrum of \kpd\ (bottom, thin gray
    line) and the FUSE spectrum (top, thin gray line) comprising a
    \ion{Si}{vii} triplet. Overplotted on each are two models (thick
    solid and dashed lines) with different Si abundances as indicated
    in the legend. Other model parameters: \Teff = 200\,000\,K, \logg
    = 6.7, C = $9.3 \times 10^{-3}$, Ne = $3.9 \times 10^{-3}$ (mass
    fractions). Observations and models were convolved with
    0.02\,\AA\ boxcars and 0.036\,\AA\ Gaussians,
    respectively.}\label{fig:si7}
\end{figure}

\section{Observations}
\label{sect:observations}

Short-slit ($0\farcs2 \times 0\farcs2$), medium-dispersion UV spectroscopy of \kpd\ in the STIS/FUV-MAMA and NUV-MAMA
configurations was performed with HST during Cycle~19, covering the
wavelength range $\sim$1150--3100\,\AA\ (Table\,\ref{tab:obs}). The
spectral resolving power as listed in Table\,\ref{tab:obs} corresponds
to $\Delta\lambda$ = 0.025--0.037\,\AA\ in the 1144--1710\,\AA\ range
and $\Delta\lambda$ = 0.054--0.102\,\AA\ in the
1616--3072\,\AA\ range. The signal-to-noise ratio (S/N) of the far-UV
spectrum below $\sim$1190\,\AA\ is relatively poor such that the
quality of the co-added archival FUSE spectra in that wavelength range is
better (see Fig.\,\ref{fig:si7} for a comparison). Identified and
unidentified photospheric lines are listed in Table\,\ref{tab:lines}.

Previous UV observations in this spectral range were performed with
the International Ultraviolet Explorer \citep[IUE,
][]{1987ApJ...321..943D} but the high-resolution spectra have
S/N that is too low to detect weak metal lines. HST data were recorded with
the Faint Object Spectrograph (FOS) with a low resolution of
2--4\,\AA\ \citep{1992ApJ...396L..79S,1994A&A...284..907W}, also
preventing the detection of weak lines. The same holds for spectra of
similar resolution taken with the Hopkins Ultraviolet Telescope
\citep[HUT,][]{1998ApJ...502..858K}. HST spectra with higher resolution were
obtained with the Goddard High Resolution Spectrograph (GHRS);
however, only a few selected, about 35\,\AA\ wide wavelength intervals
were recorded \citep{1996A&A...307..860W}.

\begin{table}
\begin{center}
\caption{Observation log of HST/STIS echelle spectroscopy.\tablefootmark{a} }
\label{tab:obs} 
\small
\begin{tabular}{ccccc}
\hline 
\hline 
\noalign{\smallskip}
Dataset & Grating & R &$\lambda$/\AA & t$_{\rm exp}$/s  \\
\hline 
\noalign{\smallskip}
OBON1010 & E230M & 30\,000 & 1616--2366 & 300 \\
OBON1020 & E230M & 30\,000 & 2277--3072 & 450 \\
OBON1030 & E140M & 45\,800 & 1144--1710 & 1215 \\
OBON1040 & E140M & 45\,800 & 1144--1710 & 3193 \\
\noalign{\smallskip} \hline
\end{tabular} 
\tablefoot{\tablefoottext{a}{Observations performed on March
    16, 2012. Spectral resolving power is R $=\lambda/\Delta\lambda$. The
    last two columns give the wavelength range covered
    and the exposure times, respectively.}} 
\end{center}
\end{table}

\begin{table}
\begin{center}
\caption{Photospheric lines detected in the HST/STIS spectra.\tablefootmark{a}}
\label{tab:lines} 
\tiny
\begin{tabular}{lccc}  
\hline 
\hline 
\noalign{\smallskip}
Wavelength/\AA  &   Ion & Transition  \\ \hline 
\noalign{\smallskip}
1162.24, 1162.67   & \ion{Ne}{viii}& 5d -- 6f \\
1164.54, 1164.76   & \ion{Ne}{viii}& 5f -- 6g \\
1164.88            & \ion{Ne}{viii}& 5g -- 6h \\
1165.94:, 1166.15:, 1166.18:&\ion{Ne}{viii}& 5f -- 6d  \\
1167.8, 1169.3     &\ion{Si}{vii} & $\rm{3s\ ^3S^o_*}- \rm{3p\ ^3P_*}$\\
1168.85, 1168.99   & \ion{C}{iv}   & 3d -- 4f \\
1171.12:, 1172.00: &\ion{O}{vi}   & 4p -- 5s  \\  
1198.55, 1198.59   & \ion{C}{iv}   & 3d -- 4p  \\ 
1207.68:           &\ion{Si}{vi}  & $\rm{3s\ ^2D_{5/2}} - \rm{3p\ ^2F^o_{7/2}}$ \\
1230.04, 1230.52   & \ion{C}{iv}   & 3p -- 4s  \\ 
1236.0             &\ion{Si}{vii} & $\rm{3s'\ ^1D^o} - \rm{3p'\ ^1F}$ \\
1238.82, 1242.80   &\ion{N}{v}    & 2s -- 2p  \\
1243.1             & ?\tablefootmark{c}             &\\
1287.80            & ?            &\\
1293.9:            &\ion{Si}{vii} & $\rm{3s\ ^1P^o}- \rm{3p\ ^1D}$\\
1315.62:, 1315.85: & \ion{C}{iv}   & 4p -- 7d \\ 
1316.3             & ?            &\\
1317.7             & ?\tablefootmark{c}             &\\
1319.78            &\ion{Ne}{vii} & $\rm{2p\ ^1P^o_1}  - \rm{2p^{2\ 3}P_2}$\\
1323.3             & ?            &\\
1351.21, 1351.29   & \ion{C}{iv}   & 4d -- 7f \\ 
1352.97            &\ion{C}{iv}   & 4f -- 7g \\
1382.1             & ?            &\\
1401.6             & ?\tablefootmark{c}             &\\
1429.1             & ?\tablefootmark{b}            & \\
1437.65            & ?\tablefootmark{c}            & \\
1440.30, 1440.38   &\ion{C}{iv}   & 4s -- 6p \\
1441.7 -- 1442.7   & ?            & \\
1450.0, 1450.55, 1451.7& ?\tablefootmark{c}             &\\
1456.8             & ?\tablefootmark{b}            & \\
1457.85, 1458.25   & ?\tablefootmark{b}            & \\
1461.5             & ?            &  \\
1480.2             & ?\tablefootmark{b}            & \\ 
1484.9             & ?\tablefootmark{c}             &\\
1548.20, 1550.77   &\ion{C}{iv}   & 2s -- 2p  \\
1548.67, 1549.34, 1549.45 & \ion{N}{v}   & 4p -- 5d  \\
1585.81, 1586.11, 1586.14 &\ion{C}{iv}   & 4p -- 6d \\
1619.62, 1619.74   & \ion{N}{v}   & 4f -- 5g & em \\
1640.42            &\ion{He}{ii}  & 2 -- 3    \\
1931.96, 1932.01, 1932.04 &\ion{Ne}{viii}& 6h -- 7i etc.& em   \\
1981.97, 1992.06, 1997.35&\ion{Ne}{vii}& $\rm{3s\ ^3S} - \rm{3p\ ^3P^o}$ \\
2070.92:, 2071.02:, 2071.06: &\ion{O}{vi}   & 5g -- 6h etc. & em  \\
2161.2:            &\ion{Ne}{vii}& $\rm{3s\ ^1P^o} - \rm{3p\ ^1D}$ & em \\
2253.39            &\ion{He}{ii}  & 3 -- 10  \\
2306.90            &\ion{He}{ii}  & 3 -- 9   \\
2386.13            &\ion{He}{ii}  & 3 -- 8   \\
2405.17, 2405.83, 2405.93 &\ion{C}{iv}   & 4p -- 5d  \\  
2511.96            &\ion{He}{ii}  & 3 -- 7   \\
2525.02, 2525.27   &\ion{C}{iv}   & 4d -- 5f  \\ 
2530.74:           &\ion{C}{iv}   & 4f -- 5g  \\ 
2698.52:, 2699.47: &\ion{C}{iv}   & 4p -- 5s \\ 
2734.11            &\ion{He}{ii}  & 3 -- 6   \\
2820.7, 2860.1     &\ion{Ne}{viii}& 3s -- 3p \\
2907.19            &\ion{C}{iv}   & 5g -- 7h   \\
2976.75            &\ion{Ne}{viii}& 7i -- 8k etc. & em  \\
2982.19            &\ion{N}{v}    & 5g -- 6h etc. & em \\
\noalign{\smallskip} \hline
\end{tabular} 
\tablefoot{
\tablefoottext{a}{``:'' denotes uncertain detection, ``?'' unidentified line, ``em'' emission line.}
\tablefoottext{b}{Also visible in archival HST spectra of the PG\,1159 stars NGC\,246 and \hh.}
\tablefoottext{c}{Also visible in \hh.}
} 
\end{center}
\end{table}

\begin{table}
\begin{center}
\caption{Number of levels and lines of model ions used for line-formation calculations of metals.\tablefootmark{a} }
\label{tab:modelatoms} 
\tiny
\begin{tabular}{cccccccc}
\hline 
\hline 
\noalign{\smallskip}
   &  IV    &   V    &    VI  &   VII  & VIII   &   IX  &    X\\ 
\hline 
\noalign{\smallskip}
C  & 54,295 &   \\   
N  &        & 27,99  & \\
O  &        & 12,16  & 54,291 \\
Ne &        &        &  8,9   & 103,761& 77,506 \\
Mg &        & 15,18  & 27,60  & 46,147 & 50,269 \\
Si &        & 25,59  & 45,193 & 61,138 & 55,239 \\
S  &        & 39,107 & 25,48  & 38,120 & 38,117 \\
Ca &        &        &        &        & 1,0    & 15,23 & 25,126\\
\noalign{\smallskip} \hline
\end{tabular} 
\tablefoot{ \tablefoottext{a}{First and second numbers of each table
    entry denote the number of levels and lines, respectively. Not listed
    for each element is the highest ionization stage considered in the 
model atom that only comprises its ground state.
    }  } 
\end{center}
\end{table}

\section{Model atoms and model atmospheres}
\label{sect:models}

For the spectral analysis we used our non-LTE
code\footnote{\url{http://astro.uni-tuebingen.de/~TMAP}} \citep{tmap2012} to compute
plane-parallel line-blanketed atmosphere models in radiative and
hydrostatic equilibrium
\citep{1999JCoAM.109...65W,2003ASPC..288...31W}. They include helium
and the four most abundant trace elements \citep[according to the
  results of][]{2010A&A...524A...9W}: C, N, O, and Ne. Four
more species (Mg, Si, S, Ca) were investigated and treated one by one
as trace elements, i.e., keeping the atmospheric structure fixed. In
the same manner, an extended model atom for Ne was introduced, meaning
that non-LTE population numbers were also computed for highly excited
levels that were treated in LTE during the preceding model-atmosphere
computations. Table\,\ref{tab:modelatoms} summarizes the number of
considered non-LTE levels and radiative transitions between them. All
model atoms were built from the publicly available T\"ubingen Model
Atom Database (TMAD\footnote{\url{http://astro.uni-tuebingen.de/~TMAD}}), comprising data
from different sources, namely \citet{1975aelg.book.....B}, the
databases of the National Institute of Standards and Technology
(NIST\footnote{\url{http://www.nist.gov/pml/data/asd.cfm}}), the
Opacity Project
\citep[OP\footnote{\url{http://cdsweb.u-strasbg.fr/topbase/topbase.html}},][]{1994MNRAS.266..805S},
CHIANTI\footnote{\url{http://www.chiantidatabase.org}}
\citep{1997A&AS..125..149D,2013ApJ...763...86L}, as well as the
Kentucky Atomic Line
List\footnote{\url{http://www.pa.uky.edu/~peter/atomic}}. 

Interstellar lines were modeled with the
program Owens \citep{2002P&SS...50.1169H,2003ApJ...599..297H}.

\section{Line identifications and spectral fitting}
\label{sect:results}

\begin{figure}[t]
 \centering  \includegraphics[width=1.0\columnwidth]{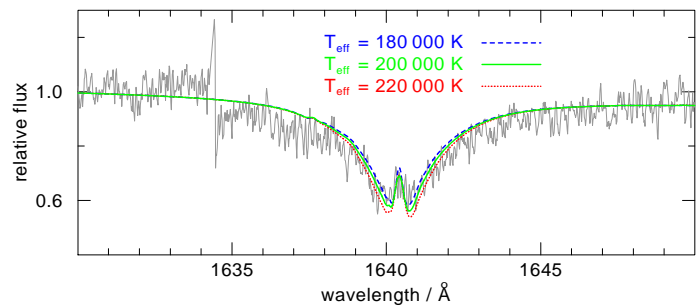}
  \caption{\ion{He}{ii} 1640\,\AA\ line in the HST/STIS spectrum (thin
    gray line). Overplotted are three models (thick lines) with
    different \Teff\ (dashed, solid, and dotted, as indicated in the
    legend) and \logg = 6.7. Observation and model were convolved with
    a 0.02\,\AA\ boxcar and a 0.036\,\AA\ Gaussian,
    respectively.}\label{fig:he}
\end{figure}

Our analysis builds on the results of \citet{2010A&A...524A...9W}. We computed a basic model with their derived values for effective
temperature (\Teff = 200\,000\,K), gravity (\logg = 6.7), and element
abundances. The model spectrum was compared to the new HST data, and
adjustments of the model parameters were made to improve the
spectral line fits.  In particular, we checked whether we can
constrain \Teff\ better than in the previous work ($\pm
20\,000$\,K). We do not improve the gravity determination of
\citet{2010A&A...524A...9W}. It was based on a careful analysis of all
\ion{He}{ii} lines in the UV and optical bands, including $\lambda
1640$\,\AA. This line is also covered by the STIS spectra
(Fig.\,\ref{fig:he}); however, the line wings are distorted because of
obvious problems with the pipeline data reduction that reveals
artificial residuals near the limits of the echelle
orders. Interestingly, however, the spectrum exhibits a weak central
emission core in the line. Close inspection of the line depth on
either side of that emission core reveals that the 200\,000\,K model
fits well and the 180\,000\,K model fits marginally better. The
220\,000\,K model can be excluded because the line profile becomes too
deep.

Generally, most spectral lines in our HST spectra are fit well by the
basic model but, as in \citet{2010A&A...524A...9W}, we encountered
problems with particular lines that are better fit with models with
slightly different parameters (\Teff, abundances). In the
following we discuss line identifications and model fits as far
as they revealed new information compared to previous work.

\begin{figure*}[bth]
 \centering  \includegraphics[width=1.0\textwidth]{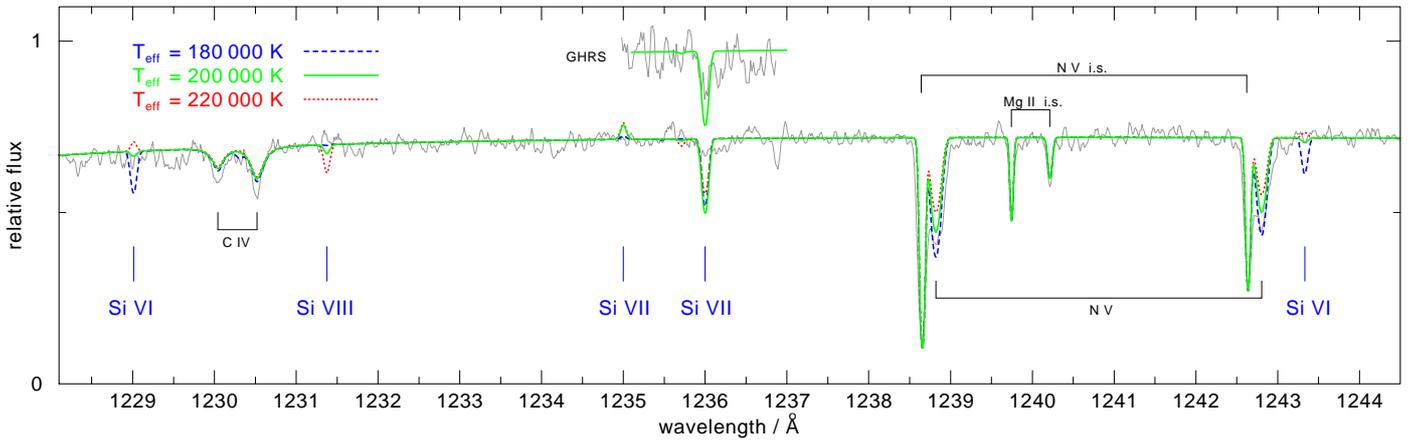}
  \caption{Section of the HST/STIS spectrum (thin gray line)
    comprising silicon lines from three ionization stages. Overplotted
    are three models (thick lines) with different \Teff\ (dashed,
    solid, and dotted, as indicated in the legend). At 180\,000\,K, the
    \ion{Si}{vi} lines are too strong; the
    \ion{Si}{viii} line is too strong at 220\,000\,K. Other model parameters: \logg =
    6.7, C = $9.3 \times 10^{-3}$, N = $2.2 \times 10^{-3}$, Si = $1.3
    \times 10^{-3}$. Also shown is the archival GHRS spectrum near the
    \ion{Si}{vii} 1236\,\AA\ line, shifted upward. Observations and
    models were convolved with 0.02\,\AA\ boxcars and
    0.036\,\AA\ Gaussians, respectively.}\label{fig:si}
\end{figure*}

\subsection{Carbon, nitrogen, and oxygen}

We see lines from \ion{C}{iv} that are well known from PG\,1159 stars,
but they are significantly weaker in \kpd\ because of the lower carbon
abundance. The abundance derived by \citet[][C =
  0.01]{2010A&A...524A...9W}\footnote{All abundances
given in mass fractions unless otherwise noted.} was a
compromise because not all \ion{C}{iv} lines in the optical and UV
spectra could be fitted with the same value. We adopted that value for
our models and kept it fixed. Two examples for lines that are too weak
in the model are those at 1169 and 1230\,\AA\ presented in
Figs.\,\ref{fig:si7} and \ref{fig:si}.

From nitrogen we see, as previously detected in HST spectra,
the \ion{N}{v} resonance doublet at 1239/1243\,\AA, blended by
stronger blueshifted interstellar components, as well as the \ion{N}{v}
5g--6h emission line at 2982.18\,\AA. We confirm the formerly
determined abundance of N = $2.5 \times 10^{-3}$.

Oxygen lines are not present. It is remarkable that the \ion{O}{v}
1371\,\AA\ line is not detected because it confirms that \Teff\ is at
least 180\,000\,K. At this temperature, the \ion{O}{vi} 5g--7h line at
1291\,\AA\ would still be detectable, and only a model with 200\,000\,K
or higher is compatible with the absence of this line in the
observation. For our modeling, we chose the abundance derived by
\citet{2010A&A...524A...9W}: O = $4 \times 10^{-3}$.

\subsection{Neon}

The presence of \ion{Ne}{viii} lines in \kpd\ was noted by
\cite{2007A&A...474..591W}. Assuming \Teff = 200\,000\,K and \logg =
7, an abundance of Ne = 1\,\% was derived from a fit to the
lines at 1162--1166\,\AA. \citet{2010A&A...524A...9W} arrived at a
slightly lower abundance of Ne = 0.4\,\% after a detailed re-analysis
that also gave a lower gravity, \logg = 6.7. Figure~\ref{fig:si7} shows a
fit to these lines with Ne = 0.39\,\% in the model.

\begin{figure}[t]
 \centering  \includegraphics[width=1.0\columnwidth]{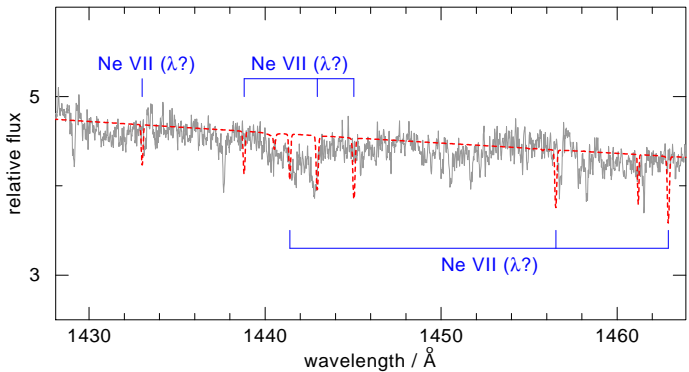}
  \caption{Section of the HST/STIS spectrum (thin gray line) and a
    model (thick dashed line) comprising two \ion{Ne}{vii} triplets
    and a singlet with uncertain wavelength positions. Some of the
    unidentified lines might correspond to the computed ones. Model
    parameters: \Teff = 200\,000\,K, \logg = 6.7, Ne = $1.2 \times
    10^{-2}$. Observation and model were convolved with a
    0.02\,\AA\ boxcar and a 0.036\,\AA\ Gaussian,
    respectively.}\label{fig:ne7_unid}
\end{figure}

\begin{figure}[bth]
 \centering  \includegraphics[width=1.0\columnwidth]{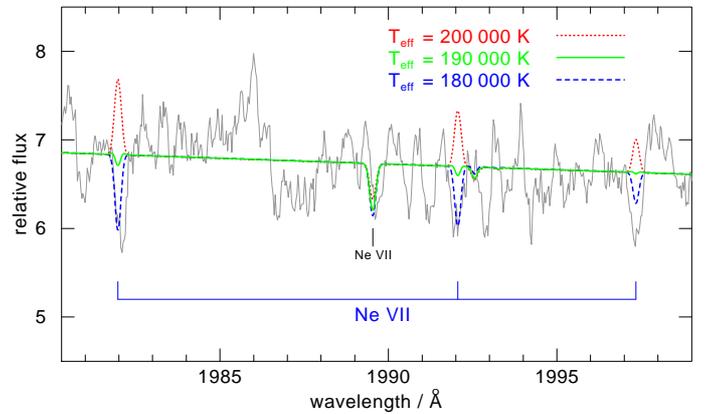}
  \caption{Spectrum detail (thin gray line) comprising a
    \Teff\ sensitive \ion{Ne}{vii} triplet. Overplotted are three
    models (thick lines) with different \Teff\ (solid, dashed, and
    dotted, as indicated in the legend). Other model parameters: \logg
    = 6.7, Ne = $1.2 \times 10^{-2}$. Observation and models were
    convolved with a 0.13\,\AA\ boxcar and 0.15\,\AA\ Gaussians,
    respectively.}\label{fig:ne7}
\end{figure}

In the new HST spectra, additional neon lines can be assessed. From
\ion{Ne}{vii}, we identify the strongest component of an
intercombination triplet at 1320\,\AA\ in absorption. Also, we see the
triplet at 1982/1992/1997\,\AA\ in absorption. A weak emission feature
near 2161\,\AA\ can be attributed to a \ion{Ne}{vii} singlet. Some
more \ion{Ne}{vii} lines are covered by the STIS spectra, but their
wavelength positions are not well known \citep[see line list in][and
  uncertainty estimates in the Kentucky
  database]{2015JPhCS.576a2007R}. Our synthetic spectra predict the
presence of such lines at positions computed from energy
levels. Because of uncertainties in the energies, the line positions
are uncertain by several \AA. Some of the unidentified lines,
therefore, could well stem from \ion{Ne}{vii}. In
Fig.\,\ref{fig:ne7_unid} we show a region where two \ion{Ne}{vii}
triplets and a singlet are located in the model and where several
unidentified lines in the observations are seen. As to \ion{Ne}{viii},
an emission feature near 1932\,\AA\ is detected, and it was previously
identified in the HST/FOS spectrum as an unresolved triplet
\citep{2007A&A...474..591W}. A rather strong \ion{Ne}{viii} doublet is
present at 2821/2861\,\AA. The following conclusions can be drawn on
\Teff\ and Ne abundance. 

\begin{figure}[bth]
 \centering  \includegraphics[width=1.0\columnwidth]{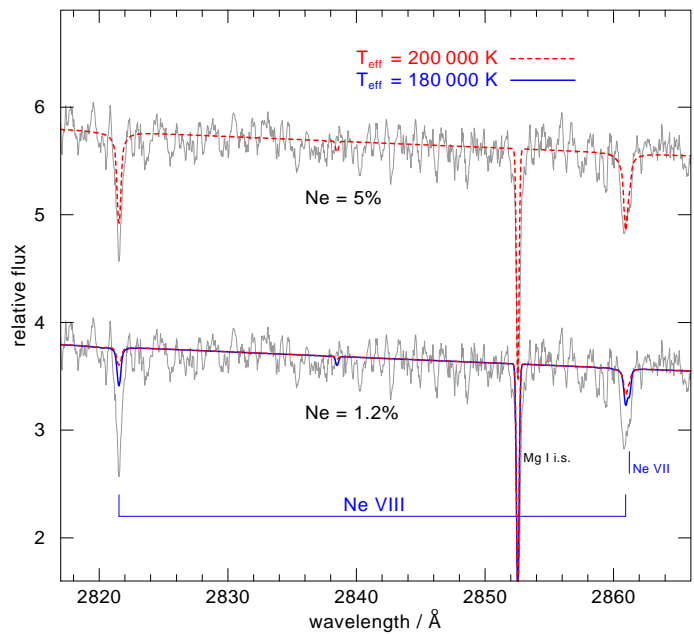}
  \caption{Detail of the NUV spectrum (thin gray lines) showing the
    \ion{Ne}{viii} 3s--3p doublet. Bottom: Overplotted are two models
    (thick lines) with different temperatures (dashed and solid lines
    as indicated in the legend) and Ne = 1.2\%. Top: Same
    spectrum and a model with Ne = 5\% (thick dashed line).
    Models have \logg = 6.7. Observation and
    models were convolved with a 0.18\,\AA\ boxcar and
    0.20\,\AA\ Gaussians, respectively.}\label{fig:ne8}
\end{figure}

With models containing Ne = 1\%, the \ion{Ne}{vii} lines fit well at
180\,000\,K, while at 200\,000\,K, the 1320\,\AA\ line is too weak, and
the 1982/1992/1997\,\AA\ triplet is in emission in contrast to weak
absorption lines seen in the observation (Fig.\,\ref{fig:ne7}). From
the \ion{Ne}{viii} lines, no clear preference for the lower or higher
\Teff\ can be derived from the 1162--1166\,\AA\ features. The
2821/2861\,\AA\ doublet poses difficulties. The lines are not deep
enough in the models. Figure\,\ref{fig:ne8} (bottom) shows that at Ne =
1\%,  neither a 180\,000\,K model nor a 200\,000\,K model fits. A
170\,000\,K model (not shown) has marginally stronger profiles. This
low temperature is, however, excluded because numerous strong
\ion{Ne}{vii} lines appear in that model that are not observed. Also,
it is at odds with the lower \Teff\ limit implied by the absence of
\ion{O}{v} 1371\,\AA.  Figure\,\ref{fig:ne8} (top) shows that an
increase in the Ne abundance to 5\% would be able to fit the
observation, but then again,  strong \ion{Ne}{vii} lines appear that
are not observed. Another problem arises with the 1932\,\AA\ line,
whose emission height is not at all achieved by any model.  In
conclusion, all the observed neon lines cannot be fit with a unique
abundance value. We adopt Ne = 1\% as a compromise.

\subsection{Magnesium}

We searched for Mg lines without success. According to our models,
\ion{Mg}{vii} has the strongest lines, but
wavelength positions are not known better than about 1--2\,\AA. At a
solar abundance level, UV lines of detectable strength are predicted,
most prominently the components of a $\rm{^3P^o} - \rm{^3P}$ triplet
at 1291--1350\,\AA. Some of the unidentified lines could therefore stem
from this ion.

\subsection{Silicon}

\ion{The Si}{vii} lines were identified in the FUSE spectrum of
\kpd,\ and Si = $1.3 \times 10^{-3}$ (twice solar) was derived
\citep{2010A&A...524A...9W}. Lines from this ion are also detectable
in the STIS spectrum, e.g., at 1167.8 and 1169.3\,\AA,\ the two
strongest components of a triplet (Fig.\,\ref{fig:si7}). A singlet at
1236.0\,\AA\ (Kentucky database wavelength with uncertainty of
0.65\,\AA) is visible in the HST/GHRS data and was already assigned to
\ion{Si}{vii} (but not fit) by \citet{2010A&A...524A...9W}. It is odd
that the line profile observed in the STIS data is much broader and
weaker than in the GHRS data (top inset in Fig.\,\ref{fig:si}) and the
computed profile. Si lines from adjacent ionization stages were not
found in the FUSE spectrum, which was only compatible with a
200\,000\,K model. The same holds for the STIS data. The 180\,000\,K
model shows prominent \ion{Si}{vi} lines, while the 220\,000\,K model
exhibits \ion{Si}{viii} lines, while both ionization stages are not
detectable in the observation (Fig.\,\ref{fig:si}). With the quoted Si
abundance, our model gives lines that are too strong: see the
\ion{Si}{vii} triplet displayed in Fig.\,\ref{fig:si7}. A good fit is
obtained with Si = $1.7 \times 10^{-4}$.

\subsection{Sulfur}

Lines of \ion{S}{vi} and \ion{S}{vii} were found in the FUSE spectrum
\citep{2010A&A...524A...9W}. No sulfur lines can be identified in the
  STIS spectra, in accordance with our model predictions.

\subsection{Calcium}

Two \ion{Ca}{x} emission lines of the 4p--4d doublet at 1136.5 and
1159.2\,\AA\ were discovered in the FUSE spectrum
\citep{2008A&A...492L..43W}. The latter is covered by our STIS
spectrum, but it cannot be identified because of the low S/N in
this region. In that paper, two absorption lines at 1461.2 and
1503.6\,\AA\ in the IUE spectrum of the hot PG\,1159 star NGC\,246
were tentatively identified as the 4s--4p doublet. No features are
seen in the STIS spectrum of \kpd\ at these wavelengths. 

In the \Teff\ range of 170\,000--220\,000\,K covered by our models, the
\ion{Ca}{x} 4s--4p lines become weaker with increasing temperature. At
solar Ca abundance (Ca = $6.4 \times 10^{-5}$), the lines become
undetectable at \Teff = 200\,000\,K, confirming the solar abundance
value derived from the 4p--4d doublet by \citet{2008A&A...492L..43W},
while models with the about four times higher value derived by
\citet{2010A&A...524A...9W} predict lines strong enough to be
detectable. The model with \Teff = 180\,000\,K and solar Ca abundance
predicts detectable lines so that at this temperature only a one-third
solar Ca abundance would explain the observation.

\subsection{Iron}

A solar abundance (Fe = $1.3 \times 10^{-3}$) was derived from
\ion{Fe}{x} lines in the FUSE spectrum
(\citet{2010A&A...524A...9W}. The star is too hot to exhibit
\ion{Fe}{vii} or \ion{Fe}{viii} lines \citep{2011A&A...531A.146W}. No
iron lines are available in the wavelength range of our STIS spectra.

\subsection{Unidentified lines}

As mentioned, some of the unidentified lines probably stem from
\ion{Ne}{vii} and \ion{Mg}{vii}. We checked the line lists cited above
for other possible candidates but found no plausible
identifications. For our search we looked for lines from light metals
heavier than CNO in ionization stages \ion{}{vi--xi} and wavelengths
known with an accuracy of at least about 0.5\,\AA\ and -- if available
-- high $gf$ values.

\begin{table}
\begin{center}
\caption{Abundances in \kpd\ ($\beta_i$) and
  in the Sun ($\beta_{\odot}$)\tablefootmark{a}. }
\label{tab:abu} 
\small
\begin{tabular}{crrrc} 
\hline 
\hline 
\noalign{\smallskip}
Element&  $\log$ $\beta_i$ & $\log$ $\beta_\odot$ & $\log(\beta_i/\beta_\odot)$ & Reference\tablefootmark{b}\\
\hline
\noalign{\smallskip} 
H  & $<-1.6$ & $-0.1$ & $<-1.5$ &  (1) \\
He & $-0.01$ & $-0.6$ & 0.6     &  (1)\\
C  & $-2.0$  & $-2.6$ & 0.6     &  (1)\\
N  & $-2.6$  & $-3.2$ & 0.6     &  (1)\\
O  & $-2.4$  & $-2.2$ & $-$0.2  &  (1)\\
Ne & $-2.0$  & $-2.9$ & 0.9     &  this work \\
Mg & $<-3.2$ & $-3.2$ & $<0.0$  &  this work \\
Si & $-3.8$  & $-3.2$ & $-$0.6  &  this work \\
S  & $-3.1$  & $-3.5$ & 0.4     &  (1)\\
Ca & $-4.2$  & $-4.2$ & 0.0     &  (2) and this work\tablefootmark{c} \\
Fe & $-2.9$  & $-2.9$ & 0.0     &  (1)\\ 
\hline
\end{tabular}
\tablefoot{ 
\tablefoottext{a}{Abundances by mass fraction. Solar abundances from \citet{2009ARA&A..47..481A}.}  
\tablefoottext{b}{References: (1) \citet{2010A&A...524A...9W}, (2) \citet{2008A&A...492L..43W}.}
\tablefoottext{c}{In this work, only an upper limit was determined.}
} 
\end{center}
\end{table}

\begin{figure*}[bth]
 \centering  \includegraphics[width=0.7\textwidth]{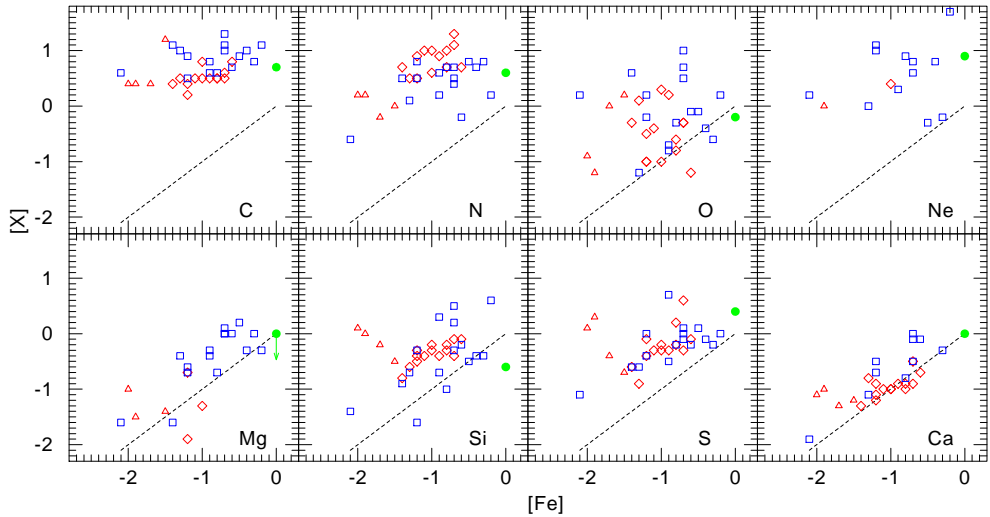}
  \caption{Observed abundances (log number relative to solar; one
    panel per species) versus iron abundance (same units) for 
    EHe stars (blue squares), majority RCB stars (red diamonds),
    minority RCB stars (red triangles) \citep[from][and references
      therein]{2011MNRAS.414.3599J} and \kpd\ (green filled
    circle). The upper limit for Mg is indicated by an arrow. The
    dashed lines indicate the solar composition scaled to
    iron.}\label{fig:abundances}
\end{figure*}

\section{Summary and discussion}
\label{sect:discussion}

The analysis of the metal lines in the new HST spectra of
\kpd\ confirms the result of \citet{2010A&A...524A...9W}, who found
\Teff = $200\,000\pm20\,000$\,K. Our investigation of the ionization
balances of neon and silicon yields a compatible result with a
slightly smaller error, namely \Teff =
$195\,000\pm15\,000$\,K. Therefore, the metal abundances determined by
\citet{2010A&A...524A...9W} are confirmed with the exception of neon
and silicon, which we improved, and an upper limit for Mg was
determined. The element abundances are summarized in
Table\,\ref{tab:abu} and displayed in Fig.\,\ref{fig:abundances},
together with results for RCB and EHe stars \citep[from][and
  references therein]{2011MNRAS.414.3599J}\footnote{Numerical
  differences between abundances of RCB and EHe stars in our
  figure and the respective Fig.\,1 in \cite{2011MNRAS.414.3599J}
  stem from our using different values for the solar
  element abundances, i.e., \citet{2009ARA&A..47..481A} instead of
  \citet{1989GeCoA..53..197A}.}. Typical errors are about 0.5\,dex for
\kpd\ and 0.3\,dex for the RCB and EHe stars. It is obvious that the
metal abundances in \kpd\ are very similar to RCB and EHe stars. 

Currently, ten objects were assigned to the group of O(He) stars
\citep{2014A&A...566A.116R,2014A&A...564A..53W,2015MNRAS.448.3587D}. They
cover a parameter range of \Teff = 80\,000\,K -- 195\,000\,K  and
\logg = 5.0 -- 6.7. (\kpd\ is the hottest member and, together with the
central star of the planetary nebula Pa\,5, has the highest
gravity.) \cite{2014A&A...566A.116R} noticed that they fall into three
subgroups: C-rich, N-rich, and C+N-rich. It was argued that
this can be explained by different scenarios within a He-WD + He-WD
merger (so-called fast-merger, slow-merger, and composite merger) for which abundance predictions were made by evolution
model calculations by \citet{2012MNRAS.426L..81Z,2012MNRAS.419..452Z}
to show that RCB stars and helium-rich subdwarf O stars could be
formed by a binary He-WD merger. Only two O(He) stars fall into the
latter group: SDSS\,J172854.34$+$361958.62 \citep[henceforth
  SDSS\,J1728;][]{2014A&A...564A..53W} and \kpd, leading
\cite{2014A&A...566A.116R} to conclude that they are the result of a
double He-WD composite merger that descended from an RCB star. With
\Teff = 100\,000\,K and \logg = 5.0, SDSS\,J1728 is located halfway
between the RCB stars and \kpd\ in the Hertzsprung Russell Diagram.

However, based on arguments from binary population synthesis, only 1\%
of the RCB stars may form from double He-WDs, and the majority forms
from a He-WD + CO-WD merger \citep[see][and references
  therein]{2014MNRAS.445..660Z}. In the cited work, it is demonstrated
that post-merger evolution calculations predict surface abundances
that can partially explain the observations in RCB stars, in
particular for the elements studied in \kpd\ in the present paper. A
comparison of the results for \kpd\ confirms earlier suggestions
\citep{2008ASPC..391..135R,2010A&A...524A...9W} that \kpd\ is indeed
an evolved RCB star. The calculations of \citet{2014MNRAS.445..660Z}
show that the observed C abundance in RCB stars can only be produced
by CO WDs in a very narrow mass range ($0.55 \pm0.02$\,M$_\odot$)
merging with a He-WD with a mass in the range
0.3--0.45\,M$_\odot$. The total mass should therefore be in the range
0.85--1.1\,$M_\odot$, which is similar to observed masses of RCB
stars, deduced from the luminosity and evolution calculations
(0.8--1.0\,M$_\odot$). The masses of the two C+N-rich O(He) stars are
$0.64\,^{+0.08}_{-0.04}$\,M$_\odot$ and
$0.73\,^{+0.14}_{-0.12}$\,M$_\odot$ for \kpd\ and SDSS\,J1728,
respectively. While the mass of SDSS\,J1728 is relatively high, that
of \kpd\ appears too low, suggesting its origin is either one of the
rare He+He WD mergers or a He+CO WD merger where the He-WD had a
significantly lower mass than the lower 0.3\,M$_\odot$ limit
postulated by \citet{2014MNRAS.445..660Z}. But we do note that the masses of
the two stars were derived with VLTP post-AGB tracks, while post-merger
tracks yield masses that are systematically higher by about
0.1--0.2\,M$_\odot$ \citep{2014A&A...566A.116R}. 

In the context of the O(He) stars it is also worth emphasizing the
finding by \cite{2014MNRAS.445..660Z} that CO+He WD mergers may also
result in stars that are \emph{not} carbon rich. Thus one could
conceive that many more of the O(He) stars are the result of such
mergers and not of He+He WD mergers, although their masses appear too low. 

To conclude, we confirm the general picture of hot helium-dominated
pre-white dwarfs as the result of binary WD mergers; however, the
detailed nature of the original binary systems remains unclear. 

\begin{acknowledgements} 
T. Rauch had been supported by the German Aerospace Center (DLR) under
grant 05\,OR\,1301. This research made use of the SIMBAD database,
operated at the CDS, Strasbourg, France, and of NASA's Astrophysics Data
System Bibliographic Services. Some of the data presented in this
paper were obtained from the Mikulski Archive for Space Telescopes
(MAST). This work used the profile fitting procedure
Owens developed by M\@. Lemoine and the FUSE French Team.
\end{acknowledgements}

\bibliographystyle{aa}  \bibliography{aa}

\end{document}